\documentstyle{l-aa}
\def \etal         {{\it et~al.}~}

\def\H2            {H{$_2$}}

\def\sec{\ifmmode{^{\prime\prime}}\else{$^{\prime\prime}$}\fi}
\def\min{\ifmmode{^{\prime}}\else{$^{\prime}$}\fi}
\def\deg{\ifmmode{^\circ}\else{$^\circ$}\fi}
\def\arcsec#1.#2 {\ifmmode {#1^{\prime\prime}\hskip-0.42em.
		  \hskip0.15em#2}
	 \else {$#1^{\prime\prime}\hskip-0.42em.\hskip0.15em#2$}
	 \fi}
\def\arcmin#1.#2 {\ifmmode {#1^{\hskip 0.05em\prime}\hskip-0.35em.
		  \hskip0.05em#2}
	 \else {$#1^{\hskip 0.05em\prime}\hskip-0.35em.\hskip0.05em#2$}
	 \fi}
\def\arcdeg#1.#2 {\ifmmode {#1\deg\hskip-0.42em.
		  \hskip0.10em#2}
	 \else {$#1\deg\hskip-0.42em.\hskip0.10em#2$}
	 \fi}

\begin{document}
\input{psfig.tex}
\thesaurus{3(11.05.2; 11.06.1; 11.09.2)  }
 
\title{ The visible environment of polar ring galaxies\thanks{based on
data obtained using the Guide Stars Selection System Astrometric
Support Program developed at the Space Telescope Science Institute}. }
\author
{Chiara Brocca\inst{1}
  \and
	Daniela Bettoni\inst{2}
  \and
	Giuseppe Galletta\inst{1}  }

\offprints{D.Bettoni}

\institute{Dipartimento di Astronomia, Universit\`a di Padova,
	Vicolo dell' Osservatorio 5, I--35122 Padova, Italy
	\and
	Osservatorio Astronomico,
	   Vicolo Osservatorio 5, I--35122 Padova, Italy. }

\date{ Received 16 December 1996; accepted 22 May 1997 }

\maketitle

\begin{abstract}
A statistical study of the environment around Polar Ring Galaxies is
presented. Two kinds of search are performed: 1) a study of the
concentration and diameters of all the objects surrounding the Polar
Rings, within a search field 5 times the ring diameter. New magnitudes
for polar ring galaxies are presented. 2) a search, in a wider field,
for galaxies of similar size that may have encountered the polar
ring host galaxy in a time of the order of 1 Gyr. Differently from
the results of similar searches in the fields of active galaxies, the
environment of the Polar Ring Galaxies seems to be similar to that of
normal galaxies.

This result may give support to the models suggesting long times for
formation and evolution of the rings. If the rings are old (and stable
or in equilibrium), no traces of the past interaction are expected in
their surroundings. In addition, the formation of massive polar rings,
too big to derive from the ingestion of a present-day dwarf galaxy,
may be easily placed in epochs with a higher number of gas-rich
galaxies. 

\end{abstract}

\section{Introduction.}\label{Intro}

The S0s with polar rings (\cite{Schweizer}) are galaxies whose
peculiarities have been originated by the accretion of matter from
outside. They show a luminous ring, composed of gas, dust and 
stars, encircling the stellar body in polar orbits. The ring may
exhibit a knotty appearance and blue colors (e.g. NGC 4650A) or a
smooth aspect and red colors (e.g. UGC 7576). The latest and widest
compilation of cases of polar ring galaxies has been made by Whitmore
\etal\ (1990).  In this catalogue (here and after called PRC) they
discuss the origin of such structures, estimating that about the 5\%
of all S0 galaxies has had a polar ring in the past, or has one now.
The external origin of the ring is explained by the fact that there is
no natural way for internal gas to set into a polar orbit and that the
large quantities of HI detected ($10^{8} \div 10^{10} M_{\odot}$) are
very unusual for early-type galaxies.

The origin and stability of polar ring galaxies are still a matter of
discussion. The current hypotheses foresee an origin linked to an
environment which should be different from that of normal
galaxies. The S0s could have ``cannibalized'' a gas-rich companion
(\cite{Quinn}, \cite{Cameron}) or could have accreted cold gas on
polar orbits from a massive disk galaxy, through a mass transfer
during a close encounter (\cite{Toomre}). Both these mechanisms are
more frequent in an environment rich in satellites or in nearby
galaxies. The PR may alternatively have accreted surrounding primeval
gas, possibly from a gas cloud (\cite{Shane}). The stability of
polar rings also shows different scenarios, as the present theories
furnish different evolutionary times: some gas-dynamics simulations
have calculated quite a fast evolution of the order of 10$^8$ yr
(\cite{Cameron}), while some N-body models foresee long formation
time-scales and slow evolution (\cite{Rix}) or even dynamical
equilibrium (\cite{Linda}, \cite{Magda}).  In the first cases the
rings now observed must be all young, unstable structures; on the contrary
if the evolutionary time scale is large or the ring is in equilibrium,
most rings, or even all of them, may be old structures.

We present here a study oriented to point out the differences, if any,
between the environment of PRs with respect to normal galaxies, in
order to discriminate among the above different hypotheses.  We
engaged two kinds of approach: i) A statistical analysis of the
objects detected in the sky region surrounding the polar ring; ii) A
survey of the galaxies with similar magnitude and red-shift as the
central object (PR or normal galaxy) in a wider region of sky.

\renewcommand{\arraystretch}[0]{0.5}
\begin{table*}
\caption[Sample]{Parameters of the sample galaxies. The polar ring 
extensions have been measured on PSS and ESO Atlases. Blue magnitudes
are from PGC-LEDA catalogue of galaxies, except for those indicated
with a symbol in the apex. The symbols represent the following
sources: + UGC; $\times$ APS; $\circ$ ROE/NRL Cosmos; * this work,
from FOCAS photometry of PDS scans. Systemic velocities V$_\odot$ are
corrected at the Sun, while Distances are calculated from corrected
distance modulus. Both data are from PGC-LEDA catalogue. The symbol
$\dag$ means that detailed photometric data for the field were not
available. }
\label{sample}
\begin{center}
\begin{tabular}{llrrlrr}
\hline
 & & & & & & \\
Name &  PRC & \multicolumn{2}{c}{Extension} & B & V$_\odot$ &  Distance \\  
 & name    &  (arcsec) & (Kpc) & magn.  & (km/s)  &  (Mpc)  \\
\hline
 & & & & & & \\
A0136-0801  & A-1 &   56   &   20  &    16.87  &      5523   &     72.8   \\
ESO415-G26  & A-2 &   54   &   15  &    14.69  &      4583   &     58.9   \\
 NGC 2685   & A-3   & 129  &  9 &  11.97 &    879  &   14.4  \\
 UGC 7576   & A-4 &   86   &   40  &    15.90  &      7036   &     95.5   \\
  NGC4650A  & A-5 &  103   &   18  &    13.91  &      2909   &     36.3   \\
 UGC 9796   & A-6   &  52  &  19 &  15.59 &   5420  &   75.2  \\
IC 51       & B-1  &   60   &   7  &    13.75  &      1758   &     24.2 \\
A0113-5442  & B-2 &   26   &   --  &    17.06$^*$  &       --   &      --   \\
  IC 1689   & B-3   &  52  &  16 &   14.8 &   4567  &   62.8  \\
A0336-4905  & B-4 &   26   &   --  &    16.19$^*$  &       --   &      --   \\
A0351-5458  & B-5 &   30   &   --  &    15.89$^*$  &       --   &      --   \\
AM0442-622  & B-6 &   34   &   --  &    16.28$^*$  &       --   &      --   \\
AM0623-371  & B-8 &   30   &   18  &    16.50$^*$  &  9745   &    127.1   \\
 UGC 5119   & B-9 &  43  &  17 &  14.55 &   5981  &   81.3  \\
 UGC 5600   & B-11  &  77  &  15 &  14.64 &   2769  &   40.4  \\
ESO503-G17$\dag$  & B-12 & 39 & -- & 16.59  &      --   &   --     \\
NGC 5122    & B-16 &   70   &  13  &    14.10  &      2939   &     38.0   \\
 UGC 9562   & B-17  &  56  &  5 &  14.38 &   1250  &   19.2  \\
AM1934-563  & B-18 &   39   &  29  &    15.97$^*$  & 11703   &    153.5   \\
AM2020-504  & B-19 &   39   &  12  &    15.21  &      4963   &      64.0   \\
ESO603-G21$\dag$  & B-21 & 52 & 11 & 15.58  &      3150   &     41.7  \\
A2329-4102  & B-22 &   47   &  --  &    15.60$^*$  &       --   &      --   \\
A2330-3751  & B-23 &   43   &  --  &    15.88$^*$  &       --   &      --   \\
A2333-1637  & B-24 &   52   &  --  &    16.20$^*$  &       --   &      --   \\
A2349-3927  & B-25 &   34   &  --  &    16.14$^*$  &       --   &      --   \\
A2350-4042  & B-26 &   26   &  --  &    16.67$^*$  &       --   &      --   \\
ESO293-G17  & B-27 &   30   &  --  &    16.17  &       --   &      --   \\
ESO349-G39  & C-1 &   82   &   --  &    15.72  &       --   &      --   \\
A0017+2212  & C-2   &  43  &  -- &  16.75$^\times$ &  --  &   --  \\
ESO474-G26  & C-3  &   32   &  33  &    14.9   &     16246   &    211.0   \\
  NGC 304   & C-6 &  64  &  21 &  13.87 &   4990  &   67.9  \\
 ESO113-G4  & C-7 &   30   &    6  &    14.96  &      3130   &     38.5   \\
ESO243-G19  & C-8 &    39   &  --  &    15.62  &       --   &      --   \\
 ESO152-G3  & C-10 &   43   &  --  &    16.72  &       --   &      --   \\
 UGC 1198   & C-12  &  39  &  4 &  15.43 &   1151  &   19.0  \\
NGC 660$\dag$ & C-13 &   138  &   8  &    11.79  &       829   &     12.0  \\
ESO199-G12  & C-15 &   86   &  36  &    15.53  &      6785   &     87.1   \\
AM0320-495  & C-16 &   19   &  --  &    16.22$^*$  &       --   &      --   \\
ESO201-G26  & C-20 &   39   &   9  &    15.13  &      3819   &     47.6   \\
A0414-4756  & C-21 &   30   &  --  &   16.0$^\circ$   &      --    &   --  \\
 ESO202-G1  & C-22 &   73   &  46  &    14.73  &     10052   &    130.6   \\
  UGC 4261  & C-24 &   56   &  24  &    14.72  &      6415   &     87.1   \\
 UGC 4323   & C-25  &  60  &  15 &  14.53 &   3691  &   52.2  \\
 UGC 4332   & C-26  &  82  &  29 &  14.83 &   5489  &   73.5  \\
NGC 2748    & C-28 &   75   &   9  &    12.4   &      1476   &     23.8  \\
NGC 2865$\dag$ & C-29 &   146  &  23  &    12.41  &      2612   &     32.7  \\
 UGC 5101   & C-30 & 60  &  48 &  15.67 &   12082  &  163.7  \\
NGC 3384$\dag$ & C-34 &   287  &  14  &    10.81  &       735   &     10.0  \\
 NGC 4174   & C-39  &  47  &  12 &  14.34 &   3813  &   52.5  \\
 UGC 7388   & C-40  &  52  &  -- &  16.00$^+$ &    --  &   --  \\
IC 3370$\dag$  & C-41 &   129  &  25  &    11.99  &      2935   &     40.2  \\
NGC 4672$\dag$ & C-42 &    86  &  17  &    14.12  &      3357   &     41.3  \\
 NGC 7468   & C-69  &  56  &  8 &  14.09 &   2085  &   29.4  \\
ZGC2315+03  & C-71  & 39  &   48 &  17.00 &   18770  &  251.2  \\
ESO240-G16$\dag$ & C-72 & 34 &  30  &    15.86  &     13664   &    179.5   \\
NGC 3718    & D-18 &   181  &  14  &    11.31  &      1031   &     16.1  \\
\hline
\end{tabular}
\end{center}
\end{table*}
\begin{table*}
\begin{tabular}{llrrlrr}
\hline
\end{tabular}
\end{table*}

\section{Selection of the sample galaxies}\label{Selection}

The widest sample of polar ring galaxies (here and after referred as
PRs) available in the literature is represented by the PRC. In this
catalogue they are divided into 4 categories. Category A is composed of 6
kinematically confirmed PRs; category B collects objects which are
good candidates for PR galaxies based on their appearance; category C
includes possible candidates and merging galaxies; finally, category D
contains an heterogeneous collection of objects such as ellipticals
with dust, boxy-bulge galaxies, etc.

As we were interested in real cases only of S0 galaxies with polar
rings, we carefully analyzed the morphology of all the galaxies in the
PRC, discarding all the doubtful or not strictly related objects. This
procedure was performed using digitized images scanned from Palomar
and ESO/SRC Sky Atlas plates.  The northern sky portion of the sample
was extracted from the data-base on optical discs of the {\it Guide
Star Selection System Astrometric Support Program} developed at the
Space Telescope Science Institute in Baltimore (STScI\footnote{STScI
is operated by the Association of Universities for Research in
Astronomy Inc., for NASA}). The PR fields of the southern sky
were obtained by scanning the glass copies of ESO/SRC J plates with
the PDS at ESO Headquarter in Garching. In both cases, the slit was 25
$\times$ 25 $\mu$, corresponding to 1.68\sec /pixel.

All the images have been inspected selecting those galaxies which
satisfied the following criteria: \\ 1) a clear presence of an
elongated structure perpendicular and external to the galaxy body; \\
2) no sign of ongoing interaction, like tails and bridges or disrupted
structures.  \\ The first point has been stated to avoid the
contamination of non-genuine PRs, such as dust-lane ellipticals or
chance superposition of far unresolved galaxies. It excludes also
small objects such as the faint Abell clusters galaxies whose
morphology is hard to distinguish in the Palomar and ESO/SRC surveys.
The second point, even if it may exclude very young or still forming
polar rings, is necessary to remove those interacting objects whose
final configuration is not expected from models to become a polar ring. We
tried, however, to discard the lowest possible number of cases,
keeping galaxies with a full ring but still interacting, such as
ESO199-G12 (C-15) but excluding the tidally interacting pair ESO566-G8 (C-31)
or structures in full merging, e.g.  NGC 7252 (D-35) and NGC 520 (D-44).

Our criteria are satisfied by the whole A-class galaxies but not by
the fainter objects of the B- and C classes. The whole D class has
been rejected but the warped galaxy NGC 3718, which has a structure
similar to that of NGC 660 but seen at a different orientation. We so
collected 56 `good' cases of PRs. These ones are listed in Table
\ref{sample}. As explained in the next Sections, in our search of
surrounding objects around polar rings we had to exclude 8 more PRs,
restricting the analysis to 48 galaxies (See Table \ref{diametri}).
This restriction was not present in our search of companions of
similar size, that has been performed on the whole sample.

We further defined a sub-set of PRs whose distances were known from
the literature. The knowledge of the distance allows us to perform a
volumetric analysis of the environment, similar to that made for other
kinds of peculiar galaxies (\cite{Dahari}, \cite{Williams},
\cite{Heckman}, \cite{Hintzen}). Unfortunately, most PRs lack known 
red-shift, and the `volume' sub-set is so restricted to 31 galaxies.

In the following sections we describe the different sources used for
obtaining the data and the methods of analysis adopted for the two
types of research.

\section{The neighborhood of the Polar Ring Galaxies}

In the first approach the extension of the search area for each field
has been established to be 5 times the diameter of the central object,
according to the previous studies on the environment of peculiar
galaxies (\cite{Theys}, \cite{Madore}).  Such a portion of sky should
be large enough to include objects able to perturb, or to have
recently perturbed, the PR host galaxy.  In our sample, the
fields have extensions of 20\min $\times$ 20\min, including much more
than five times the maximum galaxy diameter. The research area based
on diameters is advantageous because it allows to use the whole
objects sample and to refer the separation of the objects from the
central galaxy to a distance independent scale.

The normal galaxies fields, used as a control sample, were selected
using the following criteria: \\ 1) the field must be in a region of
sky as close as possible to the PR, in order to have the same
background. We used regions of equal size as the PR fields selected in
the same Schmidt plate. \\ 2) the normal galaxy must have nearly the
same apparent size as the correspondent PR.  Galaxies too big or too
small may in fact alter the counts of the background of surrounding
objects. \\

\subsection{Data production.}

The study of the environments of these fields was based on the counts
of the objects present and on the statistical analysis of their
properties, such as the projected distance $r$ from the PR and the
apparent diameter $D$. The positions and diameters within a fixed
isophote were then extracted from the APM Sky Catalogue, available
on-line from the Observatory of Edinburgh. A total of 29 fields,
mainly in the northern sky, were obtained. The data concerning two
fields selected as good examples of polar ring galaxies, ESO603-G21
and ESO503-G17, were not available. All the remaining PR fields were
analyzed from our PDS scans using the FOCAS (Faint Object
Classification and Analysis System) procedures operating in the IRAF
software package.

APM archive (\cite{APM}) furnishes data extracted from both {\em R}
and {\em B} plates of Palomar Sky Survey. It lists all the objects
present in the plates over the brightness level of 24 mag/arcsec$^2$
for the blue plates and 23 mag/arcsec$^2$ for the red plates. Their
corresponding B and R limiting magnitudes have been respectively
estimated to be 21.5 and 20.0.  The measured parameters are: the
$\alpha$ and $\delta$ coordinates, at 1950.0 equinox, the B and R
apparent magnitudes, the semi-major axis, the ellipticity and P.A. of
the ellipse fitting the image. An object is defined as non-stellar or
stellar by comparing it with the Point Spread Function of an `average'
stellar image. The objects with very small FWHM are considered as
local noise. We note that in the APM catalogue, the very large
galaxies are sometimes fragmented into many small `extended' objects
because of the identification software used.  In order to avoid this
overpopulation of false faint objects, we had to exclude these
fields. They are the regions of NGC 660, NGC 3384 and NGC 3718. This
reduced our APM sample to 24 fields. 

A different approach was needed with scans analyzed using FOCAS. The
Point Spread Function (PSF) had to be measured for each field, using
several isolated and relatively bright stars.  After that, a set of
rules was defined to classify the different kinds of object in a
similar way as the APM. After many attempts, we established that the
objects whose FWHM was between 0.6 and 1.2 times the PSF can be
considered stars; those ones between 1.21 and 10 times the PSF were
classified galaxies; while detections with smaller and larger FWHM
were considered small- and large- scale noise respectively. Here also,
three galaxies were too large for being recognized by the software as
single objects, and were fragmented into several spurious
identifications.  They are NGC 2865, NGC 4672 and IC 3370. The
corresponding fields were all discarded and the sample reduced to 24
fields.  We used the radial moments $xx$, $yy$ and $xy$, furnished by
FOCAS, to calculate the diameters $D$ in arcsec, through the formula
$$ D = 1.68
\cdot 2 \cdot \sqrt{\frac{xx+yy+\sqrt{(xx-yy)^2+4 \cdot xy^2}}{2}}.$$
We also computed the radial distance $r$ of each galaxy from the central
galaxy (PR or NG).

The final set of data includes a total of 48 PRFs (24 from APM and 24
from PDS+FOCAS). Including the control sample, the total number of
examined fields is 96.

\subsection{Magnitude calibration for PR galaxies.}

In the PRC, a lot of PRs lack B magnitude. In the automatic surveys
catalogues, such as APM or similar ones, the peculiar cross-shaped
structure of the PR often induces a false classification as ``star'',
generating unbelievably high magnitude values (from 8 to 10, in some
cases). On the contrary, the magnitudes extracted using the FOCAS
package on many galaxies of our sample were more accurate.

\begin{figure}
\psfig{file=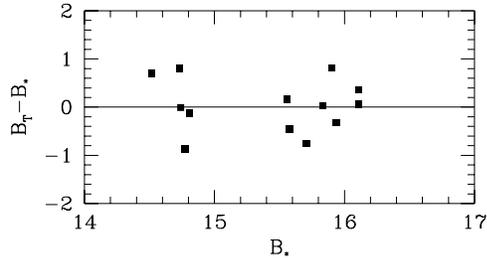,width=8.5truecm}
\caption[residuals]{Residuals for B magnitudes measured in the present work 
 and B$_T$ values from  PGC-LEDA Catalogue.}
\label{BT-BFOC}
\end{figure}

To produce new magnitude data from the scanned images, we first fixed
the zero-point level to an arbitrary sky value and then we compared
the so obtained magnitudes with those of PR galaxies whose total
magnitudes were already known. This comparison indicated a zero-point
shift of 0.71 magnitudes. When this correction was applied to the
data, the difference with the total magnitudes of the catalogues such
as RC3 (\cite{RC3}) or LEDA\footnote{The Lyon-Meudon Extragalactic
Database is supplied by the LEDA team at the CRAL-Observatoire de Lyon
(France)} became lower than half a magnitude (Figure \ref{BT-BFOC}).
The new determined magnitudes for PRs lacking this value in the
literature are listed in Table \ref{sample}.

\subsection{Statistical tests.}

\begin{figure}
\psfig{file=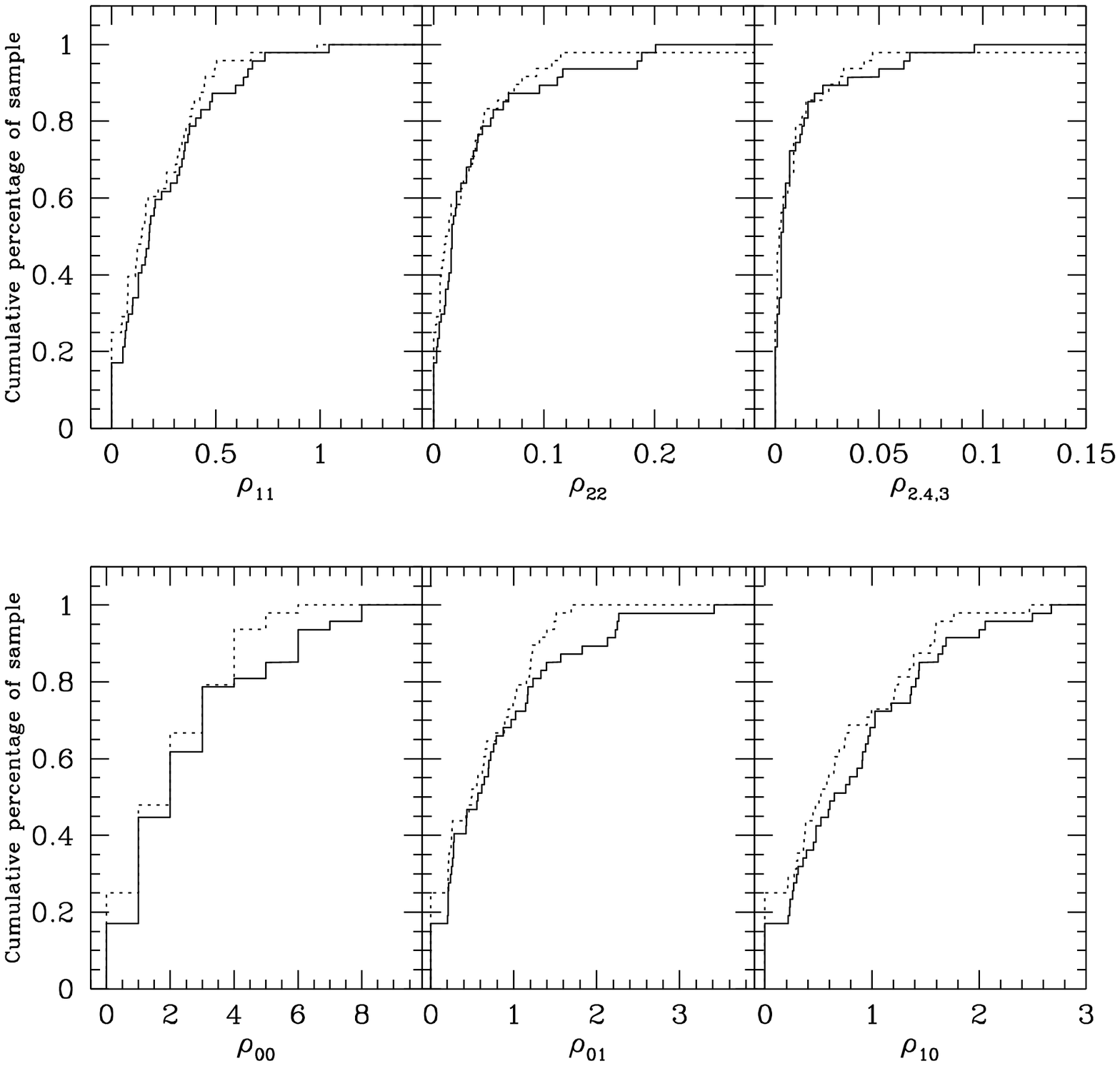,height=9truecm,angle=0}
\caption[frequency]{Cumulative frequency of $\rho_{ij}$ parameters for the
fields of polar ring galaxies (full lines) and the control fields of
normal galaxies (dotted lines). See text for the details.}
\label{Cumulative}
\end{figure}

According to similar studies (Heckman \etal 1985, Fuentes-Williams \&
Stocke 1988) we defined for each field the following density
parameters: $$ \rho_{ij}=\sum_k r_k^{-i} D_k^j, $$ where (i,j) could
assume the values 0, 1, (2,2) and (3,2.4). From the above formula,
$\rho_{00}$ represents the number of neighboring galaxies,
$\rho_{01}$ is the number weighted by the relative size, $\rho_{10}$
is weighted by proximity and $\rho_{11}$ is weighted by size and
proximity. The parameter $\rho_{22}$ is a dimensionless one
proportional to the gravitational force exerted by the surrounding
galaxies on the central object, while $\rho_{3,2.4}$ is proportional
to the tidal interaction between the surrounding galaxies and the
central one. The last two parameters were introduced by
Fuentes-Williams \& Stocke (1988). They amplify the effects present in
the parameter $\rho_{11}$.

\begin{table*}
\caption[Coefficients]{Polar rings and Normal galaxies statistics. Surrounding
objects have been searched up to 5 diameters of the central galaxy. }
\label{diametri}
\begin{center}
\begin{tabular}{lllllllllllll}
\multicolumn{7}{l}{Polar ring galaxies.} & 
\multicolumn{6}{l}{Normal galaxies.} \\
\hline
 & & & & & & & & & & & & \\
Object  & $\rho_{00}$ & $\rho_{01}$ & $\rho_{10}$ & $\rho_{11}$ & $\rho_{22}$
 & $\rho_{3,2.4}$ & $\rho_{00}$ & $\rho_{01}$ & $\rho_{10}$ & $\rho_{11}$ 
& $\rho_{22}$  & $\rho_{3,2.4}$ \\
 & & & & & & & & & & & & \\
\hline
 & & & & & & & & & & & & \\
A0136-0801 & 1. & 0.216 & 0.477 & 0.103 & 0.011 & 0.003 & 4. & 0.996 & 1.599 & 0.376 & 0.041 & 0.013 \\
ESO415-G26 & 1. & 0.203 & 0.906 & 0.184 & 0.034 & 0.016 & 0. & 0.000 & 0.000 & 0.000 & 0.000 & 0.000 \\
NGC 2685 & 1. & 0.264 & 0.257 & 0.068 & 0.005 & 0.001 & 1. & 0.425 & 0.293 & 0.125 & 0.016 & 0.003 \\
UGC 7576 & 3. & 0.971 & 1.434 & 0.471 & 0.112 & 0.062 & 4. & 1.400 & 1.205 & 0.398 & 0.043 & 0.009 \\
NGC 4650A & 2. & 0.556 & 0.646 & 0.179 & 0.016 & 0.003 & 0. & 0.000 & 0.000 & 0.000 & 0.000 & 0.000 \\
UGC 9796 & 2. & 1.171 & 0.981 & 0.594 & 0.201 & 0.096 & 2. & 0.492 & 0.471 & 0.115 & 0.007 & 0.001 \\
IC 51 & 0. & 0.000 & 0.000 & 0.000 & 0.000 & 0.000 & 0. & 0.000 & 0.000 & 0.000 & 0.000 & 0.000 \\
A0113-5442 & 2. & 0.438 & 0.940 & 0.205 & 0.021 & 0.005 & 2. & 0.499 & 0.577 & 0.143 & 0.010 & 0.002 \\
IC 1689 & 3. & 0.758 & 1.406 & 0.337 & 0.044 & 0.016 & 5. & 1.514 & 1.766 & 0.505 & 0.058 & 0.015 \\
A0336-4905 & 5. & 1.166 & 1.659 & 0.404 & 0.039 & 0.010 & 4. & 1.209 & 1.587 & 0.495 & 0.073 & 0.023 \\
A0351-5458 & 3. & 0.792 & 1.357 & 0.368 & 0.054 & 0.019 & 2. & 0.555 & 0.596 & 0.165 & 0.014 & 0.002 \\
AM0442-622 & 7. & 2.223 & 2.058 & 0.654 & 0.068 & 0.014 & 2. & 0.568 & 0.651 & 0.177 & 0.016 & 0.003 \\
AM0623-371 & 5. & 2.249 & 1.438 & 0.633 & 0.096 & 0.023 & 0. & 0.000 & 0.000 & 0.000 & 0.000 & 0.000 \\
UGC 5119 & 4. & 1.021 & 1.182 & 0.315 & 0.030 & 0.007 & 3. & 1.190 & 0.955 & 0.450 & 0.115 & 0.047 \\
UGC 5600 & 0. & 0.000 & 0.000 & 0.000 & 0.000 & 0.000 & 4. & 0.932 & 1.346 & 0.317 & 0.032 & 0.010 \\
NGC 5122 & 0. & 0.000 & 0.000 & 0.000 & 0.000 & 0.000 & 0. & 0.000 & 0.000 & 0.000 & 0.000 & 0.000 \\
UGC 9562 & 1. & 0.566 & 0.229 & 0.130 & 0.017 & 0.003 & 0. & 0.000 & 0.000 & 0.000 & 0.000 & 0.000 \\
AM1934-563 & 6. & 2.130 & 2.000 & 0.735 & 0.117 & 0.035 & 0. & 0.000 & 0.000 & 0.000 & 0.000 & 0.000 \\
AM2020-504 & 8. & 2.270 & 2.496 & 0.675 & 0.063 & 0.013 & 2. & 0.676 & 0.695 & 0.224 & 0.025 & 0.006 \\
A2329-4102 & 0. & 0.000 & 0.000 & 0.000 & 0.000 & 0.000 & 0. & 0.000 & 0.000 & 0.000 & 0.000 & 0.000 \\
A2330-3751 & 1. & 0.212 & 0.388 & 0.082 & 0.007 & 0.001 & 0. & 0.000 & 0.000 & 0.000 & 0.000 & 0.000 \\
A2333-1637 & 0. & 0.000 & 0.000 & 0.000 & 0.000 & 0.000 & 0. & 0.000 & 0.000 & 0.000 & 0.000 & 0.000 \\
A2349-3927 & 3. & 1.144 & 0.862 & 0.327 & 0.036 & 0.007 & 3. & 0.884 & 1.544 & 0.425 & 0.068 & 0.026 \\
A2350-4042 & 3. & 0.725 & 1.026 & 0.240 & 0.020 & 0.004 & 5. & 1.488 & 1.584 & 0.445 & 0.045 & 0.009 \\
ESO293-G17 & 2. & 0.424 & 0.793 & 0.167 & 0.014 & 0.003 & 0. & 0.000 & 0.000 & 0.000 & 0.000 & 0.000 \\
ESO349-G39 & 0. & 0.000 & 0.000 & 0.000 & 0.000 & 0.000 & 0. & 0.000 & 0.000 & 0.000 & 0.000 & 0.000 \\
A0017+2212 & 2. & 0.644 & 0.454 & 0.145 & 0.011 & 0.002 & 1. & 0.264 & 0.448 & 0.118 & 0.014 & 0.004 \\
ESO474-G26 & 6 & 1.566 & 1.368 & 0.353 & 0.021 & 0.003 & 3 & 0.634 & 0.99 & 0.209 & 0.16 & 0.003 \\
NGC 304 & 3. & 1.233 & 0.913 & 0.348 & 0.052 & 0.012 & 2. & 0.624 & 0.505 & 0.149 & 0.011 & 0.002 \\
ESO113-G4 & 1. & 0.254 & 0.215 & 0.055 & 0.003 & 0.000 & 3. & 0.894 & 0.995 & 0.307 & 0.038 & 0.009 \\
ESO243-G19 & 1. & 0.242 & 0.295 & 0.071 & 0.005 & 0.001 & 1. & 0.256 & 0.643 & 0.164 & 0.027 & 0.010 \\
ESO152-G3 & 3. & 0.614 & 1.026 & 0.210 & 0.016 & 0.004 & 2. & 0.660 & 0.739 & 0.265 & 0.046 & 0.015 \\
UGC 1198 & 8. & 3.418 & 2.673 & 1.043 & 0.188 & 0.050 & 3. & 0.641 & 0.748 & 0.159 & 0.009 & 0.001 \\
ESO199-G12 & 0. & 0.000 & 0.000 & 0.000 & 0.000 & 0.000 & 0. & 0.000 & 0.000 & 0.000 & 0.000 & 0.000 \\
AM0320-495 & 6. & 1.824 & 1.617 & 0.483 & 0.040 & 0.007 & 3. & 0.786 & 1.388 & 0.383 & 0.080 & 0.043 \\
ESO201-G26 & 2. & 0.429 & 0.752 & 0.162 & 0.013 & 0.003 & 1. & 0.222 & 0.358 & 0.079 & 0.006 & 0.001 \\
A0414-4756 & 1. & 0.273 & 0.474 & 0.130 & 0.017 & 0.005 & 1. & 0.243 & 0.210 & 0.051 & 0.003 & 0.000 \\
ESO202-G1 & 1. & 0.273 & 0.474 & 0.130 & 0.017 & 0.005 & 1. & 0.243 & 0.210 & 0.051 & 0.003 & 0.000 \\
UGC 4261 & 2. & 0.695 & 0.521 & 0.181 & 0.017 & 0.003 & 1. & 0.251 & 0.302 & 0.076 & 0.006 & 0.001 \\
UGC 4323 & 1. & 0.207 & 0.267 & 0.055 & 0.003 & 0.000 & 4. & 1.225 & 1.247 & 0.348 & 0.035 & 0.007 \\
UGC 4332 & 3. & 0.876 & 0.964 & 0.284 & 0.030 & 0.007 & 4. & 1.156 & 1.215 & 0.357 & 0.038 & 0.009 \\
NGC 2748 & 0. & 0.000 & 0.000 & 0.000 & 0.000 & 0.000 & 0. & 0.000 & 0.000 & 0.000 & 0.000 & 0.000 \\
UGC 5101 & 1. & 0.212 & 0.604 & 0.128 & 0.016 & 0.005 & 2. & 0.461 & 0.520 & 0.121 & 0.008 & 0.001 \\
NGC 4174 & 1. & 1.394 & 0.308 & 0.429 & 0.184 & 0.065 & 0. & 0.000 & 0.000 & 0.000 & 0.000 & 0.000 \\
UGC 7388 & 1. & 0.283 & 0.357 & 0.101 & 0.010 & 0.002 & 3. & 1.037 & 0.783 & 0.264 & 0.025 & 0.004 \\
NGC 7468 & 2. & 0.700 & 0.592 & 0.189 & 0.018 & 0.003 & 1. & 1.200 & 0.273 & 0.327 & 0.107 & 0.031 \\
ZGC2315+3 & 6. & 1.326 & 1.691 & 0.375 & 0.025 & 0.004 & 1. & 0.208 & 0.364 & 0.076 & 0.006 & 0.001 \\
ESO240-G16 & 6. & 1.693 & 2.468 & 0.669 & 0.093 & 0.033 & 1. & 0.209 & 0.374 & 0.078 & 0.006 & 0.001 \\
\hline
\end{tabular}
\end{center}
\end{table*}

The diameters and the distances from the center of the fields were
both converted in units of the central galaxy diameter and the
resulting set of parameters is listed in Table \ref{diametri}.  As
said in the previous sections, only the diffuse objects lying at 5
diameters from the center were selected, discarding those ones outside
this limit. To remove the contribution of the background galaxies, all
the objects with diameters smaller than 1/5 of the polar ring size
were excluded. Considering the real size of the galaxies with known
redshift, this cut-off limit only excludes surrounding objects with size
$\le$2-4 kpc.
 
For those polar rings whose distance is known, and for the
corresponding control sample, a set of similar parameters has been
built on a scale unit of 100 kpc and the maximum limit of the search
area has been fixed at 100 kpc from the center of the fields. The unit
and the limit assumed are similar to those used for the previous
investigations (Heckman \etal 1985) and define a research area which
is, in most cases, similar to that of the used fields (20\min). The
sample reduces to 31 objects, and the conclusions drawn are useful if
compared to those deduced from the analysis based on the
diameters. The resulting parameters are not listed here because they
bring to similar results as those of the extended sample.

\begin{table}
\caption[ks]{Summary of Kolgomorov-Smirnov tests. D$_\alpha$ is the maximum
difference observed between the two distributions, while SL is the percentage
significance level at which the two distributions compared are different. 
The negative values of D$_\alpha$ indicates a lower value of the parameter in 
the first sample with respect to the second one.}
\label{ks}
\begin{center}
\begin{tabular}{lrr}
\hline
 & & \\
 & \multicolumn{2}{c}{PR vs. NG} \\ 
 & & \\
\hline
 & & \\
Parameter & D$_\alpha$ & SL \\
    & & (\%) \\ 
 & & \\
\hline
 & & \\
\multicolumn{3}{c}{Diameter test (48 polar rings)} \\
 & & \\
 $\rho_{00}$    & -0.15 & 42\% \\
 $\rho_{01}$    & -0.15 & 41\% \\
 $\rho_{10}$    & -0.20 & 75\% \\
 $\rho_{11}$    & -0.19 & 66\% \\
 $\rho_{22}$    & -0.21 & 78\% \\
 $\rho_{2.4,3}$ & -0.23 & 85\% \\
 & & \\
\hline
 & & \\
\multicolumn{3}{c}{Volume test (31 polar rings)} \\
 & & \\
 $\rho_{00}$    & -0.14 & 10\% \\
 $\rho_{01}$    & -0.09 &  1\% \\
 $\rho_{10}$    & -0.18 & 32\% \\
 $\rho_{11}$    &  0.17 & 28\% \\
 $\rho_{22}$    & -0.19 & 37\% \\
 $\rho_{2.4,3}$ &  0.19 & 41\% \\
 & & \\
\hline
\end{tabular}
\end{center}
\end{table}

Finally, after defining the $\rho_{ij}$ parameters for the two samples
of PRs and normal galaxies scaled on diameters, a Kolgomorov-Smirnov
test has been applied by means of a Fortran program that utilizes the
IMSL library routine. The same tests have been performed for the
restricted sample of PRs for which the distance is known. The
cumulative curves are shown in Fig. \ref{Cumulative} where the fields
of PR galaxies are compared with the control fields of normal
galaxies. The results are summarized in Table \ref{ks} that will be
discussed in the Section \ref{Results}.

\section{Bright companions.}

\begin{table*}
\caption[Companions]{Possible bright companions of PR galaxies with an
estimated crossing time $<$1 Gyr. The surrounding galaxies are selected
based on the red-shift difference and separation from the PR. Their total
number is indicated in the second column (NC). For those PRs or companions
whose red-shift was not known, we indicate as possible candidates only those
placed within a 30\min radius circle and whose magnitude difference with
respect to the PR one is $\le$1. In the table, $\Delta r$ is the
separation on the sky between the PR and the nearest object.}
\label{companions}
\begin{center}
\begin{tabular}{llllll}
\hline
  &  &  &  &  & \\
 Polar     &   &  \multicolumn{4}{c}{Nearest object} \\
 Ring   &   NC & Name & $\Delta V$ & $\Delta r$ & $\Delta t_{min}$ \\
 Galaxy   &   &  & (Km/s)  &  (Kpc)     &   (Gyrs)  \\
  &  &  &  &  & \\
\hline
  &  &  &  &  & \\
A0136-0801  &  1   & PGC 6186   &   31  &  503  & 0.81 \\  
ESO415-G26  &  1   & PGC 9331   &  -53  &  302  & 0.49 \\  
  NGC 2685  &  3   & UGC 4683   &    43 &  115  & 0.19 \\  
  NGC4650A  & 16   & PGC 42951  &  -80  &   59  & 0.10 \\  
  UGC 9796  &  3   & PGC 54478  &    95 &   32  & 0.05 \\  
IC 51       &  1   & PGC0002465 &   -90 & 535   & 0.88 \\
   IC 1689  & 10   & IC 1690    &  -30  &  111  & 0.18 \\  
AM0623-371  &  3   & LEDA 96025 & -463  &  80   & 0.20 \\  
  UGC 5119  &  0   &    --      & --    & --    & --   \\  
  UGC 5600  &  4   & UGC 5609   &     5 &   14  & 0.02 \\  
ESO503-G17  &  0   &   --       &    -- &  --   &   -- \\
  UGC 9562  &  1   & UGC 9560   &   -37 &   23  & 0.04 \\  
AM1934-563  &  0   &    --      &  --   & --    & --   \\  
AM2020-504  &  2   & ESO234-G16 &  335  & 168   & 0.33 \\  
ESO603-G21  &  3   & ESO 603-G20 &   10 &  62   & 0.10 \\
   NGC 304  &  1   & UGC 591    & -148  &  433  & 0.72 \\  
 ESO113-G4  &  0   &    --      &   --  &  --   & --   \\  
NGC 660     &  7   & UGC 1195   &   -77 &  76   & 0.12 \\
  UGC 4261  &  2   & LEDA101369 &    7  &   21  & 0.03 \\  
  UGC 4323  &  1   & UGC 4376   &   416 &  421  & 0.94 \\  
  UGC 4332  &  3   & PGC 23379  &  -309 &  172  & 0.32 \\  
NGC 2748    &  1   & PGC0026654 &    15 & 260   & 0.42 \\
NGC 2865    &  0   &   --       &   --  & --    & --   \\
  UGC 5101  &  0   &    --      & --    & --    & --   \\  
NGC 3384    & 18   & NGC 3379   &   138 &  21   & 0.04 \\
  NGC 4174  &  4   & NGC 4175   &   127 &   22  & 0.04 \\  
IC 3370     &  5   & NGC4373A   &  -213 & 270   & 0.47 \\
NGC 4672    & 13   & NGC 4683   &   317 & 233   & 0.44 \\
  NGC 7468  &  8   & NGC 7454   &   -96 &  256  & 0.42 \\  
ZGC2315+03  &  0   &    --      & --    & --    & --   \\  
NGC 3718    &  8   & NGC 3729   &    19 &  55   & 0.09 \\
\hline
  &  &  &  &  & \\
Polar    &   &  \multicolumn{4}{c}{Nearest object} \\
Ring   & NC & Name & $\Delta B$ & $\Delta r$ &  \\
Galaxy &    &  & (mag.)  &  (\min)     &    \\
  &  &  &  &  & \\
\hline
  &  &  &  &  & \\
  UGC 7576  &  0   & --         &  --   &  --   & \\  
A0113-5442  &  0   &  --        &  --   & --    & \\  
A0336-4905  &  1   & PGC 13416  & -0.29 &  6.3  & \\  
A0351-5458  &  3   & ESO156-G19 &  0.37 &  3.4  & \\  
AM0442-622  &  5   & ESO84-G36  & -0.66 &  9.2  & \\  
NGC 5122    &  2   & NGC 5130   &  0.10 & 26.8  & \\
A2329-4102  &  0   &  --        &  --   & --    & \\  
A2330-3751  &  2   & LEDA 95263 & -0.48 & 22.0  & \\  
A2333-1637  &  1   & PGC 71914  & -0.70 & 18.2  & \\  
A2349-3927  & 13   & LEDA124505 &  0.73 &  5.3  & \\  
A2350-4042  &  3   & ESO293-G7  &  0.30 & 18.5  & \\  
ESO293-G17  & 17   & ESO293-G17A &  0.55 &  0.9 & \\  
ESO349-G39  &  3   & LEDA 95454 &  0.08 & 12.2  & \\  
A0017+2212  &  2   & NGC 81     &  0.99 & 19.6  & \\  
ESO474-G26  &  1   & IC 1582    &  0.60 & 12.8  & \\
ESO243-G19  &  8   & LEDA 73547 &  0.46 &  8.5  & \\  
 ESO152-G3  &  2   & ESO152-G3A & -0.90 &  0.05 & \\  
  UGC 1198  &  1   & PGC 8160   &  0.62 & 25.2  & \\  
ESO199-G12  &  1   & IC 1877    &  0.73 &  2.6  & \\  
AM0320-495  &  2   & PGC 12594  & -0.08 & 14.8  & \\  
ESO201-G26  &  1   & PGC 14720  & -0.15 & 11.7  & \\  
A0414-4756  &  2   & LEDA129487 &  0.30 & 24.3  & \\
 ESO202-G1  &  0   &   --       &  --   & --    & \\  
  UGC 7388  &  0   & --         &  --   & --    & \\  
ESO240-G16  &  1   & ESO240-G17 &  0.31 & --    & \\  
\hline
\end{tabular}
\end{center}
\end{table*}

If the accreted matter derives from a close encounter with a massive
galaxy, this galaxy may be present in the same sky region at a distance
higher than 5 PR-diameters and may have a red-shift similar to that of
the PR.  Its identification is the subject of our second analysis. We
are obliged again to divide the sample into two sets: a) 31 PRs with
known distance from red-shift, for which it is possible to define a
search volume; b) 25 remaining PRs lacking redshift, or possessing
nearby bright objects whose redshift is unknown.

\subsection{The data.}

The search has been performed on LEDA database and the results are
listed in Table \ref{companions}.  A control sample of normal galaxies
with known red-shift was also selected. The number of objects found
around every galaxy of this latter sample was then compared with that
counted for the PRs (column NC of the Table \ref{companions}).

When both the linear separation $d$ and the redshift of the companions
are known, the search area can be defined as the space that a
companion galaxy spans in a $\Delta t$ time, traveling with a relative
$\Delta V_r$ radial velocity.  The maximum projected search radius
around the PR is then $$R_s \simeq \frac{\Delta V \cdot \Delta
t}{d}.$$ We chose $\Delta V_r \cdot \Delta t =600$, adopting km/s for
the velocity and Gyr for the time. $R_s$ represents the maximum
distance covered in 1 Gyr by a galaxy moving at 600 km/s or at a
slower velocity with respect to the PR. Inside $R_s$, we selected all
the galaxies with magnitudes or redshifts similar to that of the
central PR.  For each of them, we computed, with respect to the PR,
the velocity difference $\Delta V_r$ in km/s, the linear separation
$\Delta R$ in Mpc and the minimum crossing time $\Delta t$. The
definition of these parameters is as following: $$r [kpc] =
\sqrt{\Delta \alpha['']^2 + \Delta \delta['']^2} \cdot d [Mpc]  \cdot 
(206.265)^{-1} $$ and $$\Delta t [Gyrs] = 0.97 \cdot r [kpc]/\Delta
V_{sky} [km/s] $$ having $\Delta V_{sky} = \sqrt{600^2-\Delta V_r^2}$.

With these assumptions, $\Delta t$ represents, for each companion, the
minimum time needed for this galaxy to go away from the polar ring.

For those PRs and normal galaxies without known redshift, or whose
companions lack this value, we listed only in Table \ref{companions} 
the number of galaxies present within $R_s$=30\min having a magnitude
difference $\Delta B \le$ 1 with respect to the central galaxy.

\subsection{Statistical tests.}

\begin{figure}
\psfig{file=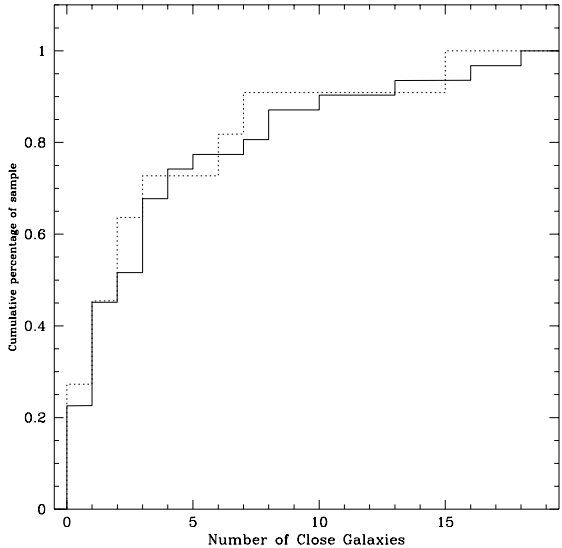,height=8truecm}
\caption[frequency2]{Cumulative frequency of the number of possible 
bright companions.  PR galaxies are represented by a solid line, while
normal galaxies data are plotted with a dashed line.}
\label{comp_ng}
\end{figure}

Examining Table \ref{companions} we note that 24 over 31 polar rings
with known distance have at least one gas-donor candidate galaxy . For
each of these PRs, we indicate in Table \ref{companions} the nearest
galaxy in terms of minimum crossing time $\Delta t_{min}$. According
to the redshift and separation from the PR, this possible donor may
have encountered the PR in a time of the order of 1 Gyr, which is
typical of the models suggesting a long evolutionary time for the
creation or the stabilization of the polar ring. A Kolgomorov-Smirnov
test has been applied to the number of possible companions found
around the PRs with respect to that found around the NGs control
sample. The cumulative distributions are shown in Figure \ref{comp_ng}.

For the remaining PRs, in the second part of Table \ref{companions}
the nearest object is listed, together with its projected separation
from the PR. Here again, 20 over 25 PRs have an object of similar
magnitude in a radius lower than 27\min. In this case, no statistical
test is possible.

\section{Results}\label{Results}

In the analysis of the PRs neighborhood, only the first three
$\rho_{ij}$ parameters are independent. They are: the number of
objects, the cumulative diameter and the objects concentration. If a
difference in the cumulative distribution of one of them exists, it
will be present and amplified by the other three $\rho_{11}$,
$\rho_{22}$, $\rho_{3,2.4}$ parameters.

The analysis of the $\rho_{ij}$ parameters shows that there are no
marked differences in the neighborhood of polar rings with respect to
that of normal galaxies. This is shown in Table \ref{ks}, where none
of the significance levels for the parameters is above 85\%. This is
still valid in the sample of PR galaxies whose distance is known,
labeled ``Volume test'' in Table \ref{ks}. In this search of close
companions, we excluded six galaxies which are too large to be
recognized as a single object by the analysis software (used by APM or
by FOCAS).  This fact may bias the sample if they would represent
extended or younger objects. However, as visible from Table
\ref{sample}, where these galaxies are identified by a $\dag$, their
linear size in Kpc is not different from that of the other PR galaxies
in the sample. If objects like NGC 660 may represents young, unstable
structures, on the other side IC 3370 or NGC 3384 have a very smooth
appearance and may be older and dynamically relaxed.

In the analysis of the bright galaxies that may have encountered the
PR before 1 Gyr, it is interesting to note the high frequency with
which at least one galaxy of similar magnitude is found in the
surrounding region. However, in comparison with the field of NGs, this
fact is not statistically significant at a level of 91\%. It is
obvious that one must be careful in applying a statistical test to an
analysis involving a single object. When a galaxy with a very similar
red-shift lies near the polar ring, it is hard to think that a
gravitational link between them is not present. The possibility that
the present bright companions of PRs interacted with them must be
analyzed for each single case.

In conclusion, the environment of PRs does not appear {\it statistically}
different from that of normal galaxies. The number of fields surveyed
in this paper is not large, but the result seems to be different from
that reached for galaxies where interaction produces an observable
nuclear activity, such as active galaxies or quasars (Dahari 1984,
Heckman \etal\ 1985, Hintzen \etal\ 1991, Rafanelli \etal 1995), whose
environments appear possibly richer than that of the normal galaxies.

Our data suggest that, if the event generating the PR was a mass
transfer from a companion galaxy or a satellite ingestion, {\it it
should have happened in a remote epoch} for the most part of galaxies
and left almost no traces in the present. This idea seems supported
by some arguments from this work and from the literature. First, the
close environment around the PRs studied does not appear perturbed at the
present epoch (Table 3). Second, some polar rings show a quantity of
gas too high to derive from the ingestion of a single dwarf, late-type
galaxy (Richter \etal\ 1994, Galletta \etal\ 1997). These massive
rings are stabilized by the mechanism of self-gravitation (Sparke
1986). Their formation may have occurred in the early phases of the
galaxy's life, when the number of late-type galaxies and their gas
content were higher and the amount of accreted gas in a single
encounter could have been large.  Third, there is at least one galaxy
of comparable size near almost all PRs (Table 4). With few exceptions
these galaxies are not at present interacting with the PR, but they
may have been gas donors in the past for the building of the
ring. Finally, some models indicate ring formation times larger than 1
Gyr (\cite{Rix}) and/or a persistence of the ring until the end of the
simulation (2.2 Gyr for \cite{Quinn} and 7.2 Gyr for
\cite{Rix}). A further support to this hypothesis may be given by the recent 
result of Reschetnikov (1997) on the detection of PRs in the Hubble
Deep Field. He found that the number of PRs present in a 5 arcmin$^2$
field, 2 objects, is consistent with a PR space density increasing in
the past.

The alternative explanations encounter some difficulties. The
hypothesis that all the PR originates from the recent accretion of small
satellites is not supported by the fact that many PRs are too massive
to derive from the gas contained in a present-day dwarf galaxy. In
addition, we may expect that an environment that is at present
favoring the formation of a PR should be different from that of normal
galaxies, which not seems confirmed by our results. The alternative that
a PR forms by means of a slow infall of diffuse, primordial gas appears
contradicted by the observations of emission lines in the rings. These
are typical of gas regions observed in our Galaxy (CO, [N II], [SII],
[OIII]) and show the presence of dust, which is a typical product of
stellar evolution.  In addition, it seems difficult for a slow infall
to produce very inclined structures of low mass, because of the tidal
torque of the host galaxy (\cite{BM}).

In conclusion, the hypothesis that the majority of present PRs are
`fossil' structures born in the early Universe may be in agreement
with the present data on the environments and with the presence of both
massive and small PRs.

\subsection*{Acknowledgments}
This work has been partially supported by the grant `Astrofisica e
Fisica Cosmica' Fondi 40\% of the Italian Ministry of University and
Scientific and Technologic Research (MURST).

\end{document}